\documentclass[letterpaper]{article}
\usepackage{iccc}
\usepackage{times}
\usepackage{helvet}
\usepackage{courier}
\usepackage{graphicx}
\usepackage{xcolor}
\usepackage{float,amsmath}
\usepackage{caption} 
\usepackage{hyperref}
\usepackage[fixed]{fontawesome5}
\title{Characterising the Creative Process in Humans and Large Language Models}
\author{{\large \bf Surabhi S. Nath$^{1,2,3}$, Peter Dayan$^{1,2}$, Claire Stevenson$^{4}$} \\
  $^1$Max Planck Institute for Biological Cybernetics, Tübingen, Germany \\
  $^2$University of Tübingen, Tübingen, Germany \\
  $^3$Max Planck School of Cognition, Leipzig, Germany \\
  $^4$University of Amsterdam, Department of Psychological Methods, Amsterdam, Netherlands
  }

\begin{document} 
\maketitle

\begin{abstract}
\begin{quote}

Large language models appear quite creative, often performing on par with the average human on creative tasks. However, research on LLM creativity has focused solely on \textit{products}, with little attention on the creative \textit{process}. Process analyses of human creativity often require hand-coded categories or exploit response times, which do not apply to LLMs. We provide an automated method to characterise how humans and LLMs explore semantic spaces on the Alternate Uses Task, and contrast with behaviour in a Verbal Fluency Task. We use sentence embeddings to identify response categories and compute semantic similarities, which we use to generate jump profiles. Our results corroborate earlier work in humans reporting both persistent (deep search in few semantic spaces) and flexible (broad search across multiple semantic spaces) pathways to creativity, where both pathways lead to similar creativity scores. LLMs were found to be biased towards either persistent or flexible paths, that varied across tasks. Though LLMs as a population match human profiles, their relationship with creativity is different, where the more flexible models score higher on creativity. Our dataset and scripts are available on \href{https://github.com/surabhisnath/Creative_Process}{GitHub}.
\end{quote}
\end{abstract}

\vspace{-0.3cm}

\section{Introduction}
Much recent work has benchmarked and quantified the generative creative aptitudes of large language models (LLMs) \cite{chakrabarty2023art,gilhooly2023ai,franceschelli2023creativity,tian2023macgyver,wang2024can,hubert2024current}. LLMs often perform as well as the average human on creative thinking tasks such as the Alternate Uses Task (AUT) \cite{orwig2024language,koivisto2023best,stevenson2022gpt3,goes2023pushing,guzik2023originality}. However, these works largely analysed creativity from a \textit{Product} perspective \cite{rhodes1961analysis}, assessing how original and useful model responses are to determine ``what makes them creative (or not)". An equally important component of creativity, less studied in the field of Artificial Creativity, is the \textit{Process} perspective \cite{rhodes1961analysis}, addressing the question of ``how creativity arises". This paper aims to fill this gap and characterise human and LLM creativity by looking at the creative process \cite{stevenson2022gpt3}, comparing the way humans and LLMs explore semantic spaces while generating sequences of creative ideas. This comparison makes sense given that the tasks involve creative response-sequence generation, in line with LLM objective of next-word prediction.

\begin{figure}
    \centering
    \includegraphics[scale=0.28]{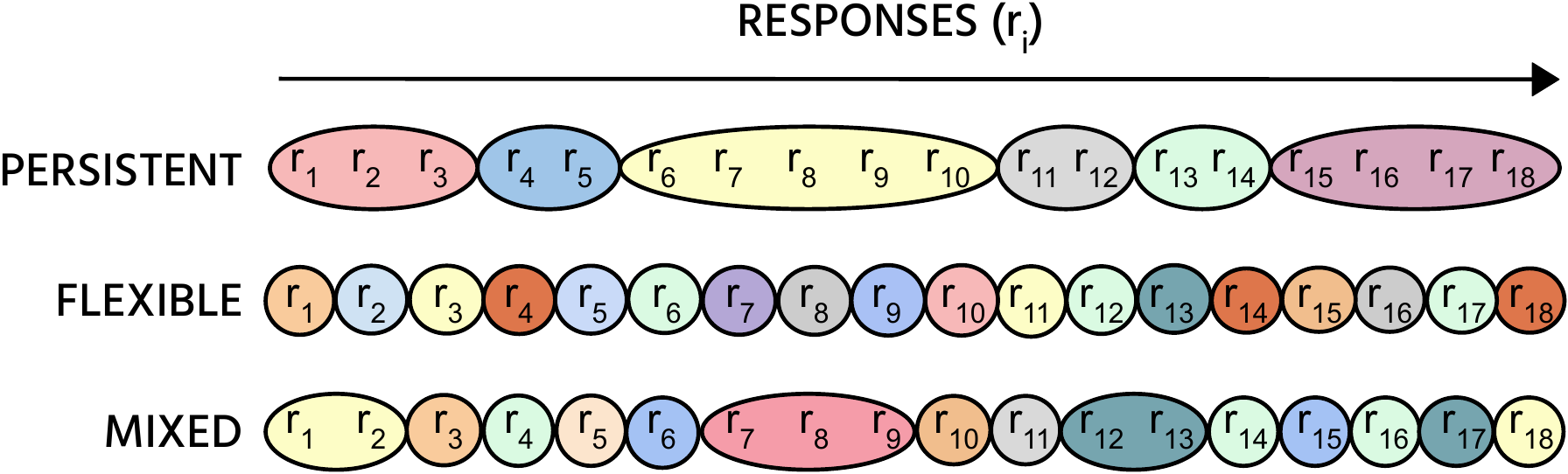}
    \caption{Example persistent, flexible and mixed response sequences. $r_{i}$ denotes the $i^{th}$ response, coloured regions denote the semantic spaces/concepts/categories. Note, in practice, most sequences will be mixed, containing different patterns of persistence and flexibility.}
    \label{fig:clusters}
    \vspace{-0.5cm}
\end{figure}

When humans generate creative ideas, for example, alternate uses for a ``brick",  two types of response pathways are observed \cite{baas2013personality,nijstad2010dual}. In the \textit{persistent} pathway,  responses stem from deeper search within limited conceptual spaces, exhibiting high clustering and similarities in responses (e.g., using a brick to break a window, break a lock, and as a nutcracker; i.e., for breaking things). In the \textit{flexible} pathway,  responses arise from broader search across multiple conceptual spaces, exhibiting frequent jumps between categories and dissimilarities in responses (e.g., using a brick to build a dollhouse, as an exercise weight, and as a coaster) (Figure \ref{fig:clusters}). 

There are two complementary ways of quantifying response clustering borrowed from the literature on memory search and semantic fluency. The first is to categorise responses temporally using inter-item retrieval times, \textit{i.e.} responses that occur shortly after each other are expected to belong to the same category and longer pauses are expected to signal jumps from one category to another. The second method is to group successive responses semantically using a set of pre-defined categories (e.g., into ``building" or ``breaking" for uses of a brick). The number of categories divided by the number of responses provides a flexibility index \cite{hills2012optimal}. Hass (\citeyear{hass2017semantic}) compared clustering in creative thinking tasks like AUT to that in a verbal fluency task (VFT) of naming animals and reported less evident clustering and higher flexibility in AUT than VFT (where responses were highly clustered, for example naming zoo animals followed by sea animals).

However, the methods used in these works are either based on handcrafted lists of categories or on response-time profiles which do not apply to responses from LLMs. In addition, these works show that semantic similarity is related to jumps in response sequences, but semantic similarity has not been used to code for jumps directly until now. In this paper, we
propose a fully automated, data-driven method to signal jumps in response sequences using response categorisation and semantic similarites and apply it to characterise the creative process in both humans and LLMs.


\begin{figure*}[h]
    \begin{minipage}{0.7\textwidth}
            \includegraphics[scale=0.3]{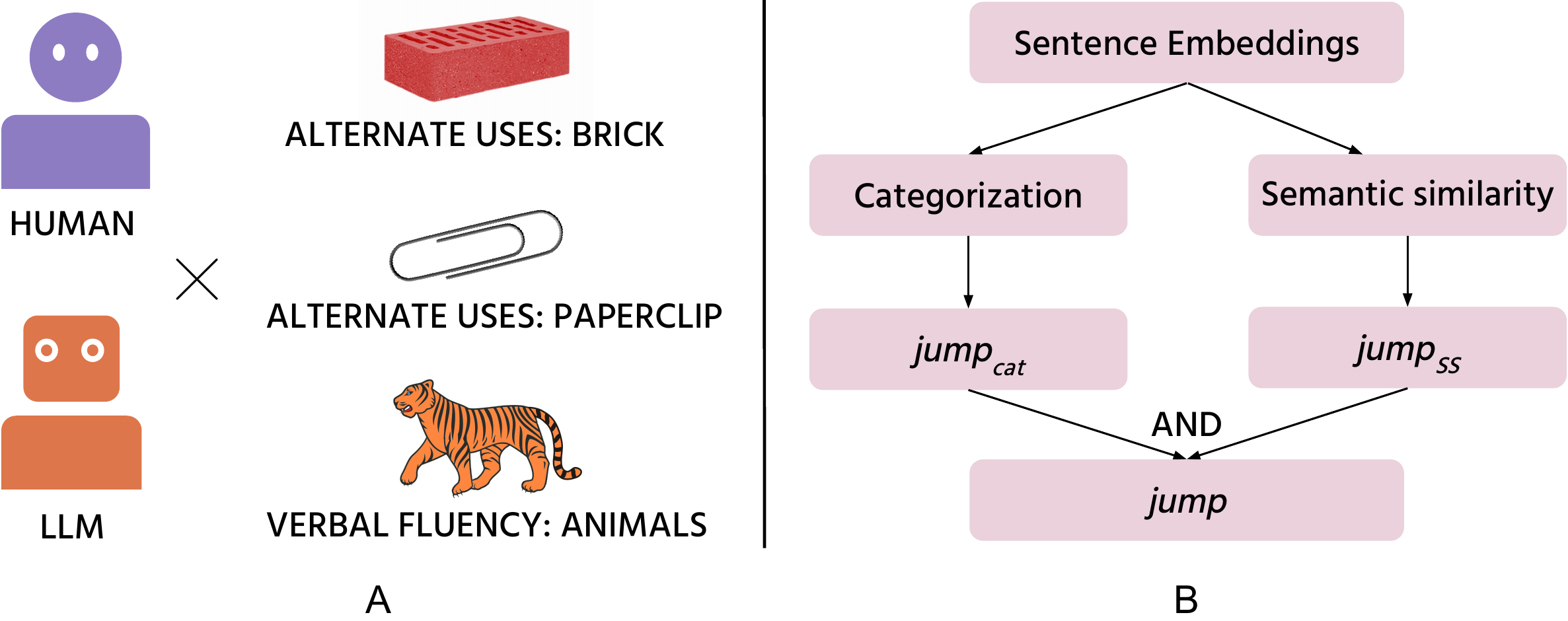} 
    \end{minipage}%
    \begin{minipage}{0.3\textwidth}
        \centering
        \caption{(A) Humans and LLMs perform 3 tasks---Alternate Uses Task (AUT) for brick and paperclip, and a Verbal Fluency Task (VFT) of naming animals. (B) Our method for obtaining jumps in the response sequence. Sentence embeddings are used for assigning response categories and evaluating semantic similarities, which respectively give \textit{jump}$_{cat}$ and \textit{jump}$_{\SS}$. Their logical AND gives \textit{jump}.}
        \label{fig:methods}
    \end{minipage}
    \vspace{-0.5cm}
\end{figure*}





\newcommand{\cat}{\textit{cat}}
\newcommand{\jump}{\textit{jump}}
\renewcommand{\SS}{\textit{SS}}


In the next sections, we first introduce our method and investigate its reliability and validity. We then apply it to characterise human and LLM flexibility on the AUT and VFT. We find that LLMs as a population match the variability in human response sequences on AUTs, but unlike humans, their relationship to creativity differs. We also discuss how to use these insights to use LLMs as artificial participants or co-creators.

\newcommand{\aut}{\text{AUT}}
\newcommand{\vft}{\text{VFT}}
\section{Method}
\paragraph{Data Collection:}
We collected data from humans and LLMs on  the AUT for ``brick" and ``paperclip", and  the VFT of naming animals (Figure \ref{fig:methods}A).

Human data were collected from undergraduate participants at University of Amsterdam using a within-subjects design. For AUT, participants listed as many creative uses for ``brick" and ``paperclip" as possible in a fixed time  of 10 minutes. For VFT, participants named as many animals as possible in a fixed time  of 2 minutes. Participants not adhering to instructions were removed, resulting in a total of 220 participants. The responses, originally in Dutch were translated to English for analysis using the \texttt{deep-translator} Python package. Translations were manually inspected to correct for errors due to spelling mistakes.

LLM data were collected in English by prompting several recent open and closed source models. For open source models, we used the Together AI API. 
The prompt matched instructions given to humans, but with specific response number and length requirements. We tested multiple prompt versions to achieve the best quality LLM responses. The final prompt for the AUT instructed LLMs to generate $n_{\aut}$ creative uses for ``brick" or ``paperclip", and to answer in short phrases of maximum $m_{\aut}$ words. For the VFT, the final prompt instructed LLMs to name $n_{\vft}$ animals, and to answer in short phrases of maximum $m_{\vft}$ words. $n_{\aut}$, $n_{\vft}$ were set to the mean number of human responses ($N$) in AUT (=ceil[max[$N_{\text{brick}}$, $N_{\text{paperclip}}$]]) and VFT tasks. $m_{\aut}$, $m_{\vft}$ were set to the maximum mean human response word length ($M$) in AUT (=floor[max[$M_{\text{brick}}$,$M_{\text{paperclip}}$]]) and VFT. 

In pilots, only $\sim$20 models gave valid responses for the AUT tasks, of which we selected the 4 that followed the prompt instructions for length and number of responses, namely, \texttt{Meta 70B Llama 3 chat HF (Llama)} model, \texttt{Mistral AI 7B Instruct (Mistral)} model, \texttt{NousResearch 7B Nous-Hermes Mistral DPO (NousResearch)} model and \texttt{Upstage 10.7B SOLAR Instruct (Upstage)} model. 

We experimented with temperature and repetition penalty parameters. However, varying the
repetition penalty did not produce higher quality responses,
so we only varied the temperature, through 11 levels (0-1, inclusive, at every 0.1). 

We also tested the latest versions of 4 closed source models: \texttt{OpenAI GPT-4 turbo (GPT)}, \texttt{Google Palm bison (Palm)}, \texttt{Google Gemini 1.0 pro (Gemini)} and \texttt{Anthropic Claude 3 (Claude)}, with the same prompt and parameters as for the open models. All 4 models generated valid responses and adhered to the response number and length instructions. 

We generated 5 samples per model $\times$ temperature combination, and therefore our LLM data set consisted of 440 (8 $\times$ 11 $\times$ 5) LLM response sequences in all.



The 220 human and 440 LLM response sequences were cleaned by removing stopwords, punctuations and common words such as ``use" or ``brick"/``paperclip". They were also manually inspected for correctness and validity. Invalid responses (verbatim repeats/junk responses) were removed \footnote{For AUT brick, we removed low temperature responses in \texttt{Mistral} and \texttt{NousResearch} models as these were verbatim repeats. For VFT, we excluded \texttt{NousResearch} and \texttt{Palm} models fully, as they only listed animals in alphabetical order.}.

\vspace{-0.3cm}

\paragraph{Response Categorisation, Semantic Similarities and Jump Signal:} 
\renewcommand{\cat}{\textit{cat}}
\renewcommand{\jump}{\textit{jump}}
\renewcommand{\SS}{\textit{SS}}
First, we encoded all responses using sentence-transformers, using the \texttt{gte-large} model given its encodings' suitability for clustering. Each response was encoded as a $1024$ dimensional normalised embedding vector. Next, all responses were aggregated, dropping duplicates, resulting in $2770$ unique alternate uses for brick, $3512$ unique alternate uses for paperclip and $482$ unique animals. The vector embeddings of these response sets were categorised using the \texttt{scipy linkage, fcluster} hierarchical clustering functions with the \texttt{ward} distance metric and a distance threshold chosen such that the mean minimum pairwise semantic similarity (vector dot product) per category was just above 0.7. This resulted in 26 brick, 28 paperclip, and 15 animal categories. 

Using these categories, we defined a binary variable $\jump_{\cat}$ for each response in a response sequence (except for the first response) as $1$ if it marked a change in category compared to the previous response, and $0$ otherwise. 

$\jump_{\cat}$ provided us course-grained similarities (for example `elevation' and `table leg' belonged to the same category as did `keep scarf together' and `hang clothes'). To address finer-grained differences, we evaluated the semantic similarity (SS) between successive embeddings of responses in a response sequence \cite{hass2017semantic,camenzind2024autoflex}. Using SS, we defined a second binary variable $\jump_{\SS}$ and set it to $0$ if SS was above a threshold and $1$ otherwise. $\jump_{\SS}$ signaled finer-grained similarities (for example`piercing' and `ring'). 

A combined jump signal was defined as their logical AND: $\jump=\jump_{\cat}\wedge\jump_{\SS}$. We set the threshold for $\jump_{\SS}$ such that  $\jump$ has at least 0.8 True Positive and True Negative Rates on hand-coded\footnote{The jump signals were hand-coded by the first author.} jump signals for AUT brick. Our entire procedure is illustrated in Figure \ref{fig:methods}B. We conduct psychometric analyses to investigate the reliability and validity of the method.

\paragraph{Jump Profiles and Participant Clustering:}
Using the jump signals, we determined a jump profile for each response sequence as the cumulative count of jumps at each response (for example, a response sequence of length 4 with jumps $[1,0,1]$ will have a jump profile $[1,1,2]$). Different human participants produced different numbers of responses, so we considered just the first 18 responses from each sequence (the median human sequence length),  excluding shorter sequences. The remaining profiles (AUT brick: 97; AUT paperclip: 103; VFT: 195) were clustered using K-Means (\texttt{sklearn KMeans}) with K-Means++ initialization~\cite{arthur2007k} per task. LLM jump profiles were assigned to the closest human cluster.

\vspace{-0.3cm}

\paragraph{Evaluating Response Creativity:}
We used Open Creativity Scoring \texttt{ocsai-chatgpt} \cite{organisciak2023beyond} to score response originality in AUT brick and paperclip.

\section{Results}

\paragraph{Jump Signal Reliability and Validity:}
We first test the reliability and validity of the $\jump$ signal. For reliability, we measured the test-retest correlation of the number of jumps for AUT brick and paperclip response sequences from 81 participants (who had ${>}{=}{18}$ responses in both). We found a positive Pearson correlation of ${r}{=}{0.42}$ (${p}{<}{0.001}$, ${CI}{=}{[0.22, 0.58]}$), which is high considering the 
test-retest and alternate-form reliability of AUT \textit{product} creativity seldom exceeds ${r}{=}{0.5}$ \cite{barbot2019measuring}. 

For validity, we test for agreement with past findings in humans. In keeping with Hass \citeyear{hass2017semantic}, who showed more jumping in AUT than VFT, we found significantly more jumps in AUT brick and paperclip than in VFT (both ${p}{<}{0.001}$). Moreover, in line with Hass (\citeyear{hass2017semantic}), Hills \textit{et al.} (\citeyear{hills2012optimal}), we also found greater mean response times for $\jump=1$ than $\jump=0$ (${p}{<}{0.001}$). 

\vspace{-0.3cm}

\begin{figure*}[ht!]
    \includegraphics[scale=0.402]
    {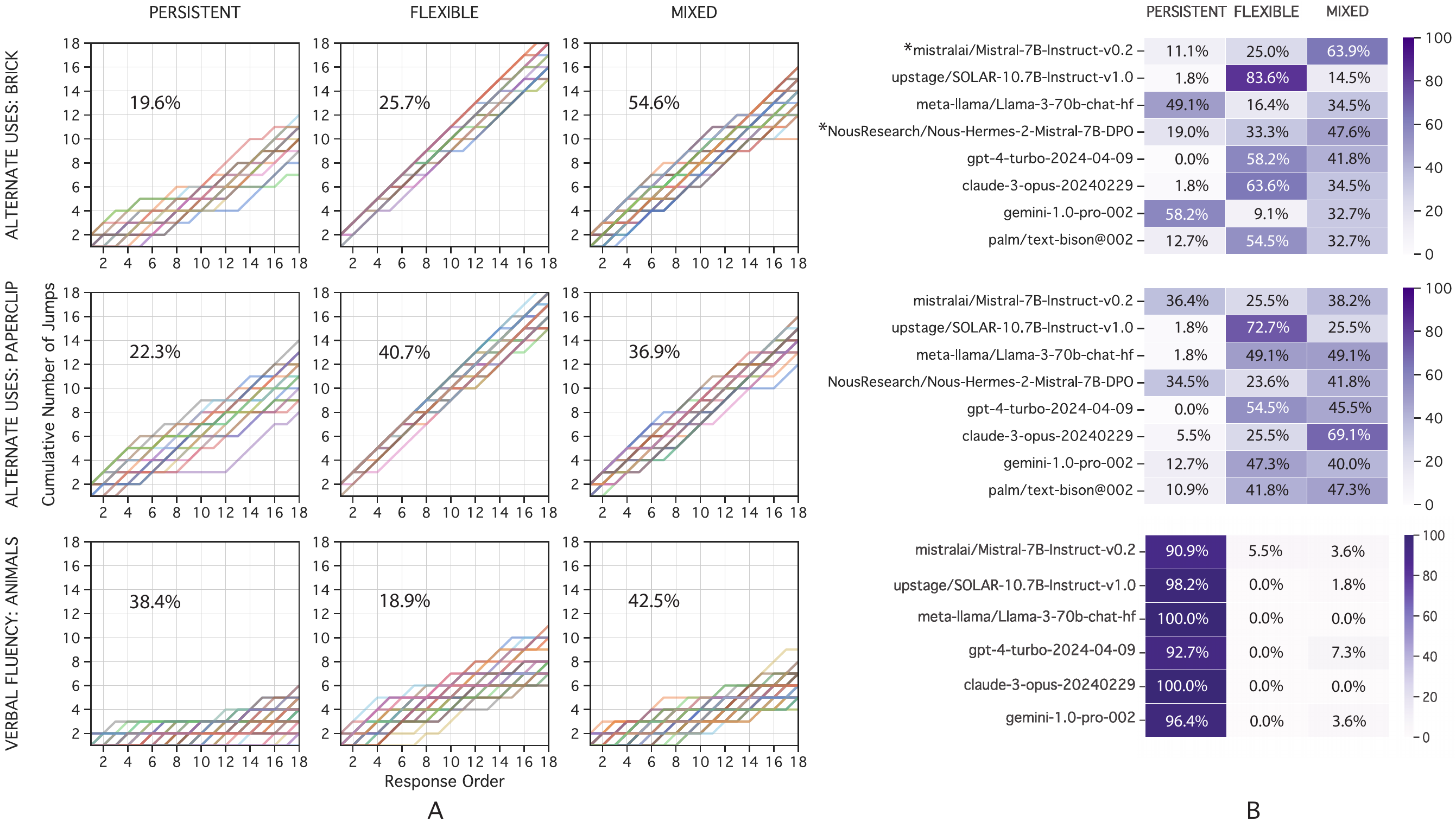}
    \vspace{-6mm}
    \caption{(A) 3 human clusters for each task--persistent, flexible and mixed. Each coloured trajectory represents 1 participant. Percentages in each row indicate the percentage of participants assigned to that cluster. (B) Percentages of each LLM response sequences assigned to each cluster. * indicates not all temperatures for that model were included (0.4-1 for \texttt{Mistral}, and 0.7-1 for \texttt{NousResearch} were used).}
    \label{fig:humanllmclusters}
    \vspace{-0.5cm}
\end{figure*}

\paragraph{Participant Clusters:}
Based on the literature and clustering elbow plots, we assigned human jump profiles to 3 clusters for each task (Figure \ref{fig:humanllmclusters}A). These map to different levels of flexibility in the response sequences---cluster 1: \textit{persistent} profiles (7-12 jumps for AUT and 1-6 jumps for VFT); cluster 2: \textit{flexible}  profiles (15-18 jumps for AUT and 6-11 jumps for VFT); and cluster 3: \textit{mixed} profiles (12-16 jumps for AUT and 4- jumps for VFT). The different numbers of jumps in AUT and VFT are clear, where the \textit{flexible} cluster in VFT closely resembles the \textit{persistent} cluster in AUTs. Thus the classifications are task-relative. The proportion of participants assigned to each cluster further reinforces that people are more flexible in AUT and more persistent in VFT.

\vspace{-0.3cm}

\paragraph{LLM Assignments:}
The LLM jump profiles were assigned to one of the 3 human clusters with proportions of assignment shown in Figure \ref{fig:humanllmclusters}B. 

Different models exhibited different biases towards persistence or flexibility in the AUTs. For example, in AUT brick,  \texttt{Upstage}, \texttt{GPT}, \texttt{Claude} and \texttt{Palm}  are mostly flexible while \texttt{Llama} and \texttt{Gemini} are mostly persistent. However, models were less consistent across the two AUTs. In AUT paperclip, while \texttt{Upstage} and \texttt{GPT} remained mostly consistent in their assignments, but \texttt{Llama} and \texttt{Gemini} switched from \textit{persistent} to \textit{flexible}. This is also evident in the test-retest correlation,  which was lower than for humans ($r{=}0.22$, ${p}{<}{0.001}$, $CI{=}[0.12, 0.31]$). Taken together, we find that LLMs are not significantly different than humans in number of jumps on AUTs (${p}{>}{0.05}$ in both). However, on the VFT, LLMs were overwhelmingly persistent, and significantly more persistent than humans (${p}{<}{0.001}$). 

Comparing the human and model cluster assignment percentages, we observe that \texttt{Mistral} and \texttt{NousResearch} models closely resemble the human distribution in AUT brick;  \texttt{Gemini} model does so in AUT paperclip; but no model resembles humans for VFT.
 
Temperature neither influenced cluster assignments nor number  of jumps in AUTs (${p}{>}{0.05}$). In VFT, temperature did influence jumping (${p}{<}{0.001}$), but did not influence cluster assignment. This is consistent with previous research suggesting no role of temperature in flexibility \cite{stevenson2022gpt3} and suggests that model responses cannot be easily manipulated parametrically.

\vspace{-0.3cm}

\paragraph{Relationship to Creativity:}
We calculated the mean originality ratings in each response sequence. For humans, mean originality was similar for \textit{persistent} and \textit{flexible} clusters in both AUTs (both ${p}{>}{0.05}$). Mean originality did not predict the number of jumps in AUT brick (${p}{>}{0.05}$), and weakly predicted jumps in AUT paperclip (${0.01}{<}{p}{<}{0.05}$). This is in line with the literature suggesting that creativity can arise both from deeper and broader search of semantic spaces.

In contrast, for LLMs, on both AUTs, mean originality was higher in the \textit{flexible} cluster compared to the \textit{persistent} cluster (both ${p}{<}{0.01}$), and mean originality predicted the number of jumps(both ${p}{<=}{0.01}$). Therefore, even though the number of jumps in AUT tasks for humans and LLMs do not differ, their relationship with originality differs. Further, LLMs also scored higher on overall mean response sequence originality compared to humans on AUTs (both ${p}{<}{0.001}$).

\section{Discussion}

We introduce an automated, data-driven method to study the creative \textit{process} in humans and LLMs. We defined an algorithmic \textit{jump signal} to indicate persistance or flexibility while solving divergent thinking tasks such as the AUT and VFT. Our jump signal proved reliable and valid for human responses and replicated findings using the traditional methods \cite{hills2012optimal,hass2017semantic}. We used this signal to investigate human and LLM jump profiles. For AUT, we found that both human and LLM jump profiles spanned from \textit{persistent} to \textit{flexible}. As in previous literature, human creativity was not correlated with flexibility profile \cite{dreu2008dual}. However, in LLMs more flexible models had higher originality scores.


Our work has limitations to be addressed with further research. First, we used the same embedding model (with different distance metrics) for response categorisation and semantic similarities. Second, each response was categorised into a single category---however, a response such as ``throw a brick to produce sound" should include concepts of both ``throw" and ``produce sound". Multiclass classification based on predefined categories could tackle this issue. Third, we used a very basic prompt for LLMs in order to match human instructions. Future work can experiment with more advanced prompt engineering. Fourth, we only scored response originality ignoring utility, which could inflate creativity comparisons as inappropriate responses (e.g., ``using a brick as a ribbon") have high originality scores but are not considered creative. Lastly, the AUT being a popular creativity task (especially AUT brick\footnote{checked with WIMBD tool \cite{elazar2023s}}) could be a source of LLM data contamination \cite{gilhooly2023ai}.

The implications of our work are many. There is an emerging trend in cognitive science to use LLMs as artificial participants \cite{argyle2023sim,frank2023llmcog,dillion2023can,binz2023turning}. Our results suggest that often LLMs are biased towards either persistence or flexibility, regardless of parameter settings such as temperature, and may be inconsistent across tasks. Therefore, we suggest using a host of models to approximate the human distribution and draw valid inferences.

The creative collaboration literature suggests that more diverse teams yield more creative ideas \cite{hoever2012diversity}.
An implication of our work for human-AI co-creativity, is to use an LLM to complement one's own brainstorming pathway. For example, more \textit{persistent} participants could collaborate with a \textit{flexible} model such as \texttt{Upstage}, which could help them diversify their ideas.
 
Through our work, we offer a first step to study human and LLM creative processes under the same metric. We provide some directions that are worth exploring in the future to further our understanding of human and artificial verbal creativity processes.

\section{Acknowledgements}
Authors thank Fleur Doorman and Emma Schreurs for their Masters theses work focussing on aspects of this project. Nath thanks Vishaal Udandarao, Andrew Webb, Sahiti Chebolu for useful suggestions and code review. The data collection and previous work on this project was funded by the Amsterdam Brain \& Cognition (ABC) Talent Grant (University of Amsterdam) 2016–2018 awarded to Stevenson and the Jacobs Foundation Fellowship 2019–2022 awarded to Stevenson (2018 1288 12).

\bibliographystyle{iccc}
\bibliography{iccc}

\end{document}